%% file: paper.tex
\documentclass{article}

\usepackage{cite,fullpage,algorithmic}

\usepackage[utf8]{inputenc} 
\usepackage[T1]{fontenc}    
\usepackage{hyperref}       
\usepackage{url}            
\usepackage{booktabs}       
\usepackage{amsfonts}       
\usepackage{nicefrac}       
\usepackage{microtype}      
\usepackage{algorithm}
\usepackage{color}
\usepackage{times}
\usepackage{graphicx}
\usepackage{subfig}

\newif\iffull
\fulltrue

\newlength\myindent
\usepackage{xspace}
\usepackage{amsthm,amsmath}
\newtheorem{theorem}{Theorem}
\newtheorem{lemma}[theorem]{Lemma}

\newtheorem{definition}[theorem]{Definition}

\newcommand{\lra}[2]{${#1}$-{\sf LRA}$_{#2}$}
\newcommand{\css}[2]{${#1}$-{\sf CSS}$_{#2}$}

\newcommand{\reals}{\mathbb{R}}
\DeclareMathOperator{\spa}{\mbox{\sl span}}
\DeclareMathOperator{\poly}{poly}

\DeclareMathOperator{\nnz}{nnz}
\newcommand{\opt}{\mathrm{opt}}

\newcommand{\comment}[1]{}

\newcommand{\mymath}[1]{$$ #1 $$}

\begin{document}

\title{Algorithms for $\ell_p$ Low-Rank Approximation}

\author{
  Flavio Chierichetti\thanks{Work done in part while visiting Google. Supported in part by a Google Focused Research Award, by the ERC Starting Grant DMAP 680153, and by the SIR Grant RBSI14Q743.}\\
Sapienza University \\
Rome, Italy \\
flavio@di.uniroma1.it
  \and
Sreenivas Gollapudi \\
Google\\
Mountain View, CA\\
sgollapu@yahoo.com
  \and
Ravi Kumar \\
Google\\
Mountain View, CA\\
ravi.k53@gmail.com
  \and
Silvio Lattanzi \\
Google\\
New York, NY\\
silviol@google.com
  \and
Rina Panigrahy \\
Google\\
Mountain View, CA\\
rinapy@gmail.com
\and
David P. Woodruff\\
IBM\\
Almaden, CA\\
dpwoodru@us.ibm.com
}
\date{}
\maketitle
\begin{abstract}
  We consider the problem of approximating a given matrix by a
  low-rank matrix so as to minimize the entrywise $\ell_p$-approximation error,
  for any $p \geq 1$; the case $p = 2$ is the classical SVD problem.
  We obtain the first provably good approximation algorithms for this
  version of low-rank approximation that work for
  every value of $p \geq 1$, including $p = \infty$. 
  Our algorithms are simple, easy to implement, work well in
  practice, and illustrate interesting tradeoffs between the
  approximation quality, the running time, and the rank of the
  approximating matrix.
\end{abstract}

\thispagestyle{empty}
\addtocounter{page}{-1}

\input intro

\input kgtone
\input iso
\input exp
\input conc
\newpage
\bibliographystyle{plain}
\bibliography{l1rpca}

\end{document}

%% file: intro.tex
\section{Introduction}
The problem of low-rank approximation of a matrix is usually studied
as approximating a given matrix by a matrix of low rank so that the
Frobenius norm of the error in the approximation is minimized.  The
Frobenius norm of a matrix is obtained by taking the sum of the
squares of the entries in the matrix. Under this objective, the
optimal solution is obtained using the singular value decomposition
(SVD) of the given matrix.  Low-rank approximation is useful in large
data analysis, especially in predicting missing entries of a matrix by
projecting the row and column entities (e.g., users and movies) into a
low-dimensional space.  In this work we consider the low-rank
approximation problem, but under the general entrywise $\ell_p$ norm, for
any $p \in [1, \infty]$.

There are several reasons for considering the $\ell_p$ version of
low-rank approximation instead of the usually studied $\ell_2$ (i.e.,
Frobenius) version.  For example, it is widely acknowledged that
the $\ell_1$ version is more robust to noise and outliers than the
$\ell_2$ version~\cite{Candes11, Huber, XY95}.  Several data mining
and computer vision-related applications exploit this insight and
resort to finding a low-rank approximation to minimize the $\ell_1$
error~\cite{lu13,meng13,wang13, Xiong11}.  Furthermore, the $\ell_1$
error is typically used as a proxy for capturing sparsity in many
applications including robust versions of PCA, sparse recovery, and
matrix completion; see, for example~\cite{Candes11, XCS10}. For these
reasons the problem has already received attention~\cite{GV15} and was
suggested as an open problem by Woodruff in his survey on sketching
techniques for linear algebra~\cite{Woodruff14}.  Likewise, the
$\ell_\infty$ version (dubbed also as the Chebyshev norm) has been
studied for the past many years~\cite{linf1, linf2}, though to the
best of our knowledge, no result with theoretical guarantees was
known for $\ell_\infty$ before our work. Our algorithm is also quite
general, and works for every $p \geq 1$. 

Working with $\ell_p$ error, however, poses many technical challenges.
First of all, unlike $\ell_2$, the general $\ell_p$ space is not
amenable to spectral techniques.  Secondly, the $\ell_p$ space is not
as nicely behaved as the $\ell_2$ space, for example, it lacks the
notion of orthogonality.  Thirdly, the $\ell_p$ version quickly runs
into computational complexity barriers: for example, even the rank-$1$
approximation in $\ell_1$ has been shown to be NP-hard by Gillis and
Vavasis~\cite{GV15}.  However, there has been no dearth in terms of
heuristics for the $\ell_p$ low-rank approximation problem, in
particular for $p =1$ and $p = \infty$: this includes alternating
convex (and, in fact, linear) minimization~\cite{KK05}, methods based
on expectation-maximization~\cite{Wetal12}, minimization with
augmented Lagrange multipliers~\cite{Zetal12}, hyperplanes projections
and linear programming~\cite{BDB13}, and generalizations of the Wiberg
algorithm~\cite{EH10}.  These heuristics, unfortunately, do not come
with any performance guarantees.  Polynomial-time algorithms for the
general problem of rank-$k$ approximation has been stated as an open
problem~\cite{Woodruff14}.  While theoretical approximation guarantees
have been given for the rank-$1$ version for the $GF(2)$ and the Boolean
cases~\cite{DHJWZ15}, to the best of our knowledge there have been no
provably good (approximation) algorithms for general matrices, or for
rank more than one, or for general $\ell_p$.

\subsection{Our Contributions}
In this paper we obtain the first provably good algorithms for the
$\ell_p$ rank-$k$ approximation problem for every $p \geq 1$.  Let $n
\times m$ be the dimensions of the input matrix.  From an algorithmic
viewpoint, there are three quantities of interest: the running time of
the algorithm, the approximation factor guaranteed by the algorithm,
and the actual number of vectors in the low-rank approximation that is
output by the algorithm (even though we only desire $k$). 

Given this setting, we show three main algorithmic results intended
for the case when $k$ is not too large.  First, we show that one can
obtain a $(k+1)$-approximation to the rank-$k$ problem in time $m^k
\poly(n,m)$; note that this running time is not polynomial once $k$ is
larger than a constant.  To address this, next we show that one can get an
$O(k)$-approximation to the best $k$-factorization in time
$O(\poly(nm))$; however, the algorithm returns $O(k \log m)$ columns,
which is more than the desired $k$ (this is referred to as a
bi-criteria approximation).  Finally, we combine these two algorithms.
We first show that the output of the second algorithm can further be
refined to output exactly $k$ vectors, with an approximation factor of
$\poly(k)$ and a running time of $O(\poly(n,m)(k \log n)^k )$.  The
running time now is polynomial as long as $k = O(\log n / \log \log n)$.
Next, we show that for any constant $p \geq 1$,
we can obtain approximation factor
$(k \log m)^{O(p)}$ and a running time of $\poly(n,m)$ for every
value of $k$. 

Our first algorithm is existential in nature: it shows that there are
$k$ columns in the given matrix that can be used, along with an
appropriate convex program, to obtain a $(k+1)$-approximation.  Realizing
this as an algorithm would therefore na\"ively incur a factor $m^k$ in the
running time.  Our second algorithm works by sampling columns and
iteratively ``covering'' the columns of the given matrix, for an
appropriate notion of covering. In each round of sampling our algorithm
uniformly samples from a remaining set of columns; we note here that it
is critical that our algorithm is adaptive as otherwise uniform sampling
would not work. While this is computationally efficient
and maintains an $O(k)$-approximation to the best rank-$k$ approximation,
it can end up with more than $k$ columns, in fact $O(k \log m)$.
Our third algorithm fixes
this issue by combining the first algorithm with the notion of a 
near-isoperimetric transformation for the $\ell_p$-space, which lets
us transform a given matrix into another matrix spanning the same
subspace but with small $\ell_p$ distortion.

A useful feature of our algorithms is that they are uniform with
respect to all values of $p$.  We test the performance of our
algorithms, for $p = 1$ and $p = \infty$, on real and synthetic data
and show that they produce low-rank approximations that are
substantially better than what the SVD (i.e., $p = 2$) would obtain.


\subsection{Related Work}
In \cite{Woodruff2016},
a low-rank approximation was obtained which holds for every $p \in [1, 2]$.
Their main
result is an $(O(\log m) \poly(k))$-approximation in
$\nnz(A) +(n+m)\poly(k)$ time, for every $k$,
where $\nnz(A)$ is the number of non-zero entries
in $A$. In our work, we
also obtain such a result for $p \in [1,2]$ via very
different sampling-based methods, whereas the results in
\cite{Woodruff2016} are sketching-based.
In addition to that, we obtain
an algorithm with a $\poly(k)$ approximation factor which is independent
of $m$ and $n$, though this latter algorithm requires
$k = O(\log n / \log \log n)$ in order to be polynomial time. 

Another result in
~\cite{Woodruff2016} shows how to achieve a $k \poly(\log k)$-approximation,
in $n^{O(k)}$ time for $p \in [1,2]$.
For $k$ larger than a constant, this is larger
than polynomial time, whereas our algorithm with $\poly(k)$
approximation ratio is polynomial time for 
$k$ as large as $\Theta(\log n / \log \log n)$.

Importantly, our results
hold for every $p \geq 1$, rather than only $p \in [1,2]$,
so for example, include $p = \infty$. 

In addition we note that there exist papers solving problems that, at
first blush, might seem similar to ours. For instance,
\cite{deshpande11} study a convex relation, and a rounding algorithm
to solve the subspace approximation problem (an $\ell_p$
generalization of the least squares fit), which is related to but
different from our problem.  Also, \cite{feldman07} offer a
bi-criteria solution for another related problem of approximating a
set of points by a collection of flats; they use convex relaxations to
solve their problem and are limited to bi-criteria solutions, unlike
ours. Finally, in some special settings robust PCA can be used to
solve $\ell_1$ low-rank approximation~\cite{Candes11}.  However,
robust PCA and $\ell_1$ low-rank approximation have some apparent
similarities but they have key differences.  Firstly, $\ell_1$
low-rank approximation allows to recover an approximating matrix of
any chosen rank, whereas RPCA returns some matrix of some unknown
(possibly full) rank. While variants of robust PCA have been proposed
to force the output rank to be a given value \cite{nnsa014,yi16}, these variants
make additional noise model and incoherence assumptions
on the input matrix, whereas our results
hold for every input matrix.
Secondly, in terms of approximation quality, it
is unclear if near-optimal solutions of robust PCA provide near-optimal
solutions for $\ell_1$ low-rank approximation. 

Finally, we mention concrete example matrices $A$ for which the SVD
gives a poor approximation factor for $\ell_p$-approximation error. First,
suppose $p < 2$ and $k = 1$. Consider the following $n \times n$ block
diagonal matrix composed of two blocks: a $1 \times 1$ matrix with
value $n$ and an $(n-1) \times (n-1)$ matrix with all $1$s. The SVD
returns as a solution the first column, and therefore incurs polynomial
in $n$ error for $p = 2 - \Omega(1)$. Now suppose $p > 2$ and $k = 1$. Consider
the following $n \times n$ block diagonal matrix composed of two
blocks: a $1 \times 1$ matrix with value $n-2$ and an $(n-1) \times (n-1)$
matrix with all $1$s. The SVD returns as a solution the matrix
spanned by the bottom block, and so also incurs an error polynomial
in $n$ for $p = 2 + \Omega(1)$. 

\section{Background}

For a matrix $M$, let $M_{i,j}$ denote the entry in its $i$th row and
$j$th column and let $M_i$ denote its $i$th column.  Let $M^T$ denote
its transpose and let $|M|_p = \left(\sum_{i, j}
  |M_{i,j}|^p\right)^{1/p}$ denote its entry-wise $p$ norm.
Given a set $S = \{ i_1, \ldots, i_t \}$ of column indices, let
$M_S = M_{i_1,\ldots, i_t}$ be the matrix composed of the columns of
$M$ with the indices in $S$.

Given a matrix $M$ with $m$ columns, we will use $\spa M =
\left\{\sum_{i = 1}^m \alpha_i M_i \mid \alpha_i \in \reals \right\}$
to denote the vectors spanned by its columns.  If $M$ is a matrix and
$v$ is a vector, we let $d_p(v, M)$ denote the minimum $\ell_p$
distance between $v$ and a vector in $\spa M$: \mymath{ d_p(v, M) =
  \inf_{w \in \spa M} |v - w|_p.  }

Let $A \in \reals^{n \times m}$ denote the input matrix and let $k > 0$
denote the target rank.  We assume, without loss of generality (w.l.o.g.),
that $m \le n$. Our first
goal is, given $A$ and $k$, to find a subset $U \in \reals^{n \times k}$
of $k$ columns of $A$ and $V \in \reals^{k \times m}$ so as to minimize the
$\ell_p$ error, $p \geq 1$, given by 
\mymath{
|A - U V|_p.
}

Our second goal is, given $A$ and $k$, to find $U \in \reals^{n \times
  k}, V \in \reals^{k \times m}$ to minimize the $\ell_p$ error, $p
\geq 1$, given by
\mymath{
|A - U V|_p.
}
Note that in the second goal, we do not require $U$ be a subset of columns.

We refer to the first problem as the \emph{$k$-columns subset
  selection problem in the $\ell_p$ norm}, denoted \css{k}{p}, and to
the second problem as the \emph{rank-$k$ approximation problem in the
  $\ell_p$ norm}, denoted \lra{k}{p}.%
\footnote{\lra{k}{2} is the classical SVD problem of finding $U \in
  \reals^{n \times k}, V \in \reals^{k \times m}$ so as to minimize
  $|A-UV|_2$.}  In the paper we often call $U, V$ the
\emph{$k$-factorization} of $A$.  Note that a solution to \css{k}{p}
can be used as a solution to \lra{k}{p}, but not necessarily
vice-versa.

In this paper we focus on solving the two problems for general $p$.
Let $U^{\star} V^{\star}$ be a $k$-factorization of $A$ that is
optimal in the $\ell_p$ norm, where $U^{\star} \in \reals^{n \times
  k}$ and $V^{\star} \in \reals^{k \times m}$, and let
$\text{opt}_{k,p}(A) = |A - U^{\star} V^{\star}|_p$.  An algorithm is
said to be an \emph{$\alpha$-approximation}, for an $\alpha \geq 1$,
if it outputs $U \in
\reals^{n \times k}, V \in \reals^{k \times m}$ such that \mymath{ |A
  - U V|_p \leq
\alpha \cdot \opt_{k,p}(A).
}

It is often convenient to view the input matrix as $A = U^{\star}
V^{\star} + \Delta = A^{\star} + \Delta$, where $\Delta$ is some
\emph{error} matrix of minimum $\ell_p$-norm.  Let $\delta =
|\Delta|_p = \opt_{k,p}(A)$.

We will use the following observation.
\begin{lemma}\label{lem:secondlpfactor}
  Let $U \in \reals^{n \times k}$ and $v \in \reals^{n \times 1}$.
  Suppose that there exists $x \in \reals^{k \times 1}$ such that
  $\delta = \left|U \cdot x - v\right|_p$.  Then, there exists a polynomial
  time algorithm that, given $U$ and $v$, finds $y \in \reals^{k \times 1}$
  such that $\left|U \cdot y - v\right|_p \le \delta$.
\end{lemma}
\begin{proof}
  This $\ell_p$ regression problem is a convex program and well-known
  to be solvable in polynomial time.
\end{proof}

%% file: kgtone.tex
\newcommand{\myand}{\mbox{ {\rm and } } }
\newcommand{\SelectColumns}{{\sc SelectColumns}\xspace}

\section{An $(m^{k} \poly(nm))$-time algorithm for \lra{k}{p}}
\label{sec:alg1}

In this section we will present an algorithm that runs in time $m^{k}
\poly(nm)$ and produces a $(k+1)$-approximation to \css{k}{p} (so also
to \lra{k}{p}) of a matrix $A \in \reals^{n \times m}$, $m \le n$, for
any $p \in [1, \infty]$.  The algorithm simply tries all possible
subsets of $k$ columns of $A$ for producing one of the factors, $U$,
and then uses Lemma~\ref{lem:secondlpfactor} to find the second factor
$V$.

\subsection{The existence of one factor in $A$}

For simplicity, we assume that $|\Delta_i|_p > 0$ for each column $i$.
To satisfy this, we can add an arbitrary small random error to each
entry of the matrix.  For instance, for any $\gamma > 0$, and to
each entry of the matrix, we can add an independent uniform value in
$[-\gamma,\gamma]$. This would guarantee that $|\Delta_i|_p > 0$
for each $i \in [m]$.

Recall that $A = A^{\star} + \Delta$ is the perturbed matrix, and we
only have access to $A$, not $A^{\star}$.  Consider
Algorithm~\ref{alg:existential} and its output $S$. Note that we cannot
actually run this algorithm since we do not know $A^{\star}$. Thus,
it is a hypothetical algorithm used for the purpose of our proof.
That is, the 
algorithm serves as a proof that there exists a subset of $k$ columns
of $A$ providing a good low rank approximation. In
Theorem~\ref{thm:kcol} we prove that the columns in $A$ indexed by the
subset $S$ can be used as one factor of a $k$-factorization of $A$.

\begin{algorithm}[h]
\begin{algorithmic}[1]
\REQUIRE A rank $k$ matrix $A^{\star}$ and perturbation matrix $\Delta$ 
\ENSURE $k$ column indices of $A^{\star}$
\STATE For each column index $i$, let $\tilde{A}^{\star}_{i} \leftarrow A^{\star}_{i} / |\Delta_{i}|_p$.
\STATE Write $\tilde{A}^{\star} = \tilde{U} \cdot \tilde{V}$, s.t.
$\tilde{U} \in \mathbb{R}^{n \times k}, \tilde{V} \in \mathbb{R}^{k \times m}$. 
\STATE Let $S$ be the subset of $k$ columns of $\tilde{V} \in \mathbb{R}^{k \times m}$
that has maximum determinant in absolute value (note that the subset $S$ indexes
a $k \times k$ submatrix). 
\STATE Output $S$.
\end{algorithmic}
\caption{\label{alg:existential}
  Enumerating and selecting $k$ columns
  of $A$.}  
\end{algorithm}

Before proving the main theorem of the subsection, we show a useful property of
the matrix $\tilde{A}^{\star}$, that is, the matrix having the vector 
$A^{\star}_{i} / |\Delta_{i}|_p$ as the $i$th column. Then we will
use this property to prove Theorem~\ref{thm:kcol}.

\begin{lemma}\label{lem:smallCoefficients}
  For each column $\tilde{A}^\star_i$ of $\tilde{A}^\star$, one can write
  $\tilde{A}^\star_i = \sum_{j \in S} M_{i}(j) \tilde{A}^\star_j$,
  where $|M_{i}(j)| \leq 1$ for all $i, j$.
  \end{lemma}
\begin{proof}
  Fix an $i \in \{1, \ldots, m\}$.  Consider the equation $\tilde{V}_S
  M_i = \tilde{V}_i$ for $M_i \in \mathbb{R}^k$.  We can assume the
  columns in $\tilde{V}_S$ are linearly independent, since w.l.o.g.,
  $\tilde{A}^\star$ has rank $k$. Hence, there is a unique solution $M_i =
  (\tilde{V}_S)^{-1} \tilde{V}_i$. By Cramer's rule, the $j$th
  coordinate $M_{i}(j)$ of $M_i$ satisfies $M_i(j) =
  \frac{\det(\tilde{V}^j_S)}{\det(\tilde{V}_S)}$, where
  $\tilde{V}^j_S$ is the matrix obtained by replacing the $j$th
  column of $\tilde{V}_S$ with $\tilde{V}_i$. By our choice of $S$,
  $|\det(\tilde{V}^j_S)| \leq |\det(\tilde{V}_S)|$, which implies
  $|M_{i}(j)| \leq 1$.  Multiplying both sides of equation
  $\tilde{V}_S M_i = \tilde{V}_i$ by $\tilde{U}$, we have
  $\tilde{A}^{\star}_SM_i = \tilde{A}^{\star}_i$.
  \end{proof}
  
  Now we prove the main theorem of this subsection.

\begin{theorem}\label{thm:kcol}
  Let $U = A_S$.  For $p \in [1, \infty]$, let $M_1, \ldots, M_m$ be
  the vectors whose existence is guaranteed by
  Lemma~\ref{lem:smallCoefficients} and let $V \in \reals^{k \times n}$
  be the matrix having the vector $|\Delta_i|_p \cdot \left( M_i(1) /
    |\Delta_{i_1}|_p, \ldots, M_i(k) / |\Delta_{i_k}|_p \right)^T$ as
  its $i$th column. Then, $|A_i - (U V)_i|_p \le (k+1) |\Delta_i|_p$
  and hence $|A - U V|_p \le (k+1) |\Delta|_p$.
\end{theorem}

\begin{proof}
We consider the generic column $(UV)_i$. 
\begin{eqnarray*}
\lefteqn{
  (UV)_i \enspace = \enspace |\Delta_i|_p  \sum_{j=1}^k
  \left(\frac{M_i(j)}{|\Delta_{i_j}|_p}  A_{i_j}\right)} \\
&=& |\Delta_i|_p  \sum_{j=1}^k \left(\frac{M_i(j)}{|\Delta_{i_j}|_p}  \left(A^{\star}_{i_j} + \Delta_{i_j}\right)\right)
  \\
&=& |\Delta_i|_p  \sum_{j=1}^k \left(M_i(j)  \tilde{A}^{\star}_{i_j} +
  M_i(j)  \frac{\Delta_{i_j}}{|\Delta_{i_j}|_p}\right) \\
&=& |\Delta_i|_p  \tilde{A}^{\star}_i + |\Delta_i|_p  \sum_{j=1}^k \left( M_i(j)  \frac{\Delta_{i_j}}{|\Delta_{i_j}|_p}\right) \\
&=& A^{\star}_i + \sum_{j=1}^k \left( |\Delta_i|_p  \cdot M_i(j)
  \frac{\Delta_{i_j}}{|\Delta_{i_j}|_p}\right)
\stackrel{\Delta}{=} A^{\star}_i + E_i.
\end{eqnarray*}

Observe that $E_i$ is the weighted sum of $k$ vectors,
$\frac{\Delta_{i_1}}{|\Delta_{i_1}|_p}, \ldots,
\frac{\Delta_{i_k}}{|\Delta_{i_k}|_p}$, having unit $\ell_p$-norm.
Observe further that, since the sum of their weights satisfies
$|\Delta_i|_p \sum_{j=1}^k |M_i(j)| \le k|\Delta_i|_p$, we have the
$\ell_p$-norm of $E_i$ is not larger than $|E_i|_p \le k
|\Delta_i|_p$.  The proof is complete using the triangle inequality:
\begin{eqnarray*}
|A_i - (UV)_i|_p & \le & |A^{\star}_i - A_i|_p + |A^{\star}_i - (UV)_i|_p \\
& = & |\Delta_i|_p + |E_i|_p \\
& \leq & (k + 1) |\Delta_i|_p.
\quad
\qedhere
\end{eqnarray*}
\end{proof}

\subsection{An $m^{k} \poly(nm)$-time algorithm}
In this section we give an algorithm that returns a
$(k+1)$-approximation to the \lra{k}{p} problem in time $m^{k}
\poly(nm)$.
\begin{algorithm}[h]
\begin{algorithmic}[1]
  \REQUIRE An integer $k$ and a matrix $A$ 
  \ENSURE $U \in \reals^{n
    \times k}, V \in \reals^{k \times m}$ s.t. $|A - U V|_p \le (k+1)
  \opt_{k,p}(A)$.  
  \FORALL{$I \in \binom{[m]}k$} 
  \STATE Let $U = A_I$
  \STATE Use Lemma~\ref{lem:secondlpfactor} to compute a matrix $V$
  that minimizes the distance $d_I = |A - U V|_p$
\ENDFOR
\STATE Return $U, V$ that minimizes $d_I$, for $I \in \binom{[m]}k$
\end{algorithmic}
\caption{\label{alg:nk} A $(k+1)$-approximation to \lra{k}{p}.}
\end{algorithm}

The following statement follows directly from the existence of $k$
columns in $A$ that make up a factor $U$ having small $\ell_p$ error
(Theorem~\ref{thm:kcol}).

\begin{theorem}\label{thm:nk}
  Algorithm~\ref{alg:nk} obtains a $(k+1)$-approximation to 
\lra{k}{p} in time $m^{k} \poly(nm)$.
\end{theorem}

\section{A $\poly(nm)$-time bi-criteria algorithm for \css{k}{p}}
\label{sec:alg2}

We next show an algorithm that runs in time $\poly(nm)$ but returns
$O(k \log m)$ columns of $A$ that can be used in place of $U$, with an
error $O(k)$ times the error of the best $k$-factorization.  In other
words, it obtains more than $k$ columns but achieves a polynomial
running time; we will later build upon this algorithm in
Section~\ref{sec:alg3} to obtain a faster algorithm for the \lra{k}{p}
problem.  We also show a lower bound: there exists a matrix $A$ for
which the best possible approximation for the \css{k}{p}, for $p\in
(2,\infty)$, is $k^{\Omega(1)}$.

\begin{definition}[Approximate coverage]
  Let $S$ be a subset of $k$ column indices. We say that column $A_i$ is
  {\bf $c_p$-approximately covered} by $S$ if for $p \in [1, \infty)$ we
  have $\min_{x\in  \reals^{k \times 1}} |A_Sx-A_i|_p^p \leq c
  \frac{100(k+1)^p|\Delta|_p^p}{n}$, and for $p = \infty$, $\min_{x\in  \reals^{k \times 1}}
  |A_Sx-A_i|_{\infty} \leq c(k+1)|\Delta|_{\infty}$.  If $c = 1$, we
  say $A_i$ is {\bf covered} by $S$.
\end{definition}
We first show that if we select a set $R$ columns of size $2k$ uniformly at
random in $\binom{[m]}{2k}$, with constant probability we cover a
constant fraction of columns of $A$.
\newcommand{\CE}{\mathcal{E}}
\begin{lemma}
\label{lem:cover}
Suppose $R$ is a set of $2k$ uniformly random chosen columns of
$A$. With probability at least $2/9$, $R$ covers at least a $1/10$
fraction of columns of $A$.
\end{lemma}
\begin{proof}
  Let $i$ be a column index of $A$ selected uniformly at
  random and not in the set $R$.
  Let $T = R \cup \{i\}$ and let $\eta$ be the cost of the
  best $\ell_p$ rank-$k$ approximation to $A_T$. Note that $T$
  is a uniformly random subset of $2k+1$ columns of $A$. 

  {\bf Case: $p < \infty$.} Since
  $T$ is a uniformly random subset of $2k+1$ columns of $A$,
  ${\bf E}_T[\eta^p] = \frac{(2k+1)|\Delta|_p^p}{n}$. 
  Let $\CE_1$ denote the event ``$\eta^p \leq \frac{10(2k+1)|\Delta|_p^p}{n}$''.
  By a Markov bound, $\Pr[\CE_1] \geq 9/10$.

  By Theorem \ref{thm:kcol}, there exists a subset $L$ of $k$ columns
  of $A_T$ for which $\min_X |A_LX-A_T|_p^p \leq (k+1)^p
  \eta^p$. Since $i$ is itself uniformly random in the set $T$, it
  holds that ${\bf E}_i[\min_x |A_Lx-A_i|_p^p \leq \frac{(k+1)^p
    \eta^p}{2k+1}$.  Let $\CE_2$ denote the event ``$\min_x
  |A_Lx-A_i|_p^p \leq \frac{10(k+1)^p\eta^p}{2k+1}$''.  By a Markov
  bound, $\Pr[\CE_2] \geq 9/10$.

  Let $\CE_3$ denote the event ``$i \notin L$''.  Since $i$ is
  uniformly random in the set $T$, $\Pr[\CE_3] \geq \frac{k+1}{2k} >
  1/2$. 
    
  Clearly $\Pr[\CE_1 \wedge \CE_2 \wedge \CE_3] \geq 3/10$.
  Conditioned on $\CE_1 \wedge \CE_2 \wedge \CE_3$, we have
\begin{eqnarray*}
\min_x |A_Rx-A_i|_p^p & \stackrel{\CE_3}{\leq} & \min_x |A_Lx-A_i|_p^p \\
& \stackrel{\CE_2}{\leq} & \frac{10(k+1)^p \eta^p}{2k+1} \\
& \stackrel{\CE_1}{\leq} & \frac{100(k+1)^p|\Delta|_p^p}{n},
\end{eqnarray*}
which implies that $i$ is covered by $R$. Note that the first inequality
uses that $L$ is a subset of $R$ given $\mathcal{E}_3$,
and so the regression cost using $A_L$
cannot be smaller than that of using $A_R$
    
Let $Z_i$ be an indicator variable if $i$ is covered by $R$ and let $Z
= \sum_i Z_i$.  We have ${\bf E}[Z] = \sum_i {\bf E}[Z_i] \geq \sum_i
\frac{3}{10} = 3m/10$; hence ${\bf E}[m-Z] \leq \frac{7m}{10}$.  By
a Markov bound, $\Pr[m-Z \geq \frac{9m}{10}] \leq \frac{7}{9}$.

{\bf Case $p = \infty$.} Then $\eta \leq |\Delta|_{\infty}$ since
$A_T$ is a submatrix of $A$. By Theorem \ref{thm:kcol}, there exists a
subset $L$ of $k$ columns of $A_T$ for which $\min_X
|A_LX-A_T|_{\infty} \leq (k+1)\eta$.  Defining $\CE_3$ as before and
conditioning on it, we have
\begin{eqnarray*}
\min_x |A_Rx-A_i|_{\infty} & \leq & \min_x |A_Lx-A_i|_{\infty} \\
& \leq & \min_X |A_LX-A_T|_{\infty} \\
& \leq & (k+1) |\Delta|_{\infty},
\end{eqnarray*}
i.e., $i$ is covered by $R$.  Again defining $Z_i$ to be the event
that $i$ is covered by $R$, we have ${\bf E}[Z_i] \geq \frac{1}{2}$,
and so ${\bf E}[m-Z] \leq \frac{m}{2}$, which implies $\Pr[m-Z \geq
\frac{9m}{10}] \leq \frac{5}{9} < \frac{7}{9}$.
\end{proof}

We are now ready to introduce Algorithm~\ref{alg:randomcolumns}.  We
can without loss of generality assume that
the algorithm knows a number $N$ for which
$|\Delta|_p \leq N \leq 2|\Delta|_p$.
Indeed, such a value can be obtained by first computing
$|\Delta|_2$ using the SVD. Note that although one does not know
$\Delta$, one does know $|\Delta|_2$ since this is the Euclidean
norm of all but the top $k$ singular values of $A$, which one can
compute from the SVD of $A$. Then, note that for $p < 2$, $|\Delta|_2
\leq |\Delta|_p \leq n^{2-p}|\Delta|_2$, while for $p \geq 2$,
$|\Delta|_p \leq |\Delta|_2 \leq n^{1-2/p}|\Delta|_p$. Hence, there
are only $O(\log n)$ values of $N$ to try, given $|\Delta|_2$, one of
which will satisfy $|\Delta|_p \leq N \leq 2|\Delta|_p$. One can take
the best solution found by Algorithm~\ref{alg:randomcolumns} for each
of the $O(\log n)$ guesses to $N$.
\begin{algorithm}[h]
\begin{algorithmic}
  \REQUIRE An integer $k$, and a matrix $A = A^{\star} + \Delta$.
\ENSURE $O(k \log m)$ columns of $A$ 
\STATE \hspace*{-5mm} \SelectColumns($k, A$)
\IF{number of columns of $A \le 2k$}
\STATE return all the columns of $A$
\ELSE
\REPEAT
\STATE Let $R$ be uniform at random $2k$ columns of $A$
\UNTIL {at least $(1/10)$-fraction columns of $A$ are $c_p$-approximately covered}
\STATE Let $A_{\overline{R}}$ be the columns of $A$ not approximately
covered by $R$
\STATE return $A_R \cup$ \SelectColumns($k, A_{\overline{R}}$)
\ENDIF
\end{algorithmic}
  \caption{\label{alg:randomcolumns} Selecting $O(k \log m)$ columns of $A$.}
\end{algorithm}
\begin{theorem}
  With probability at least $9/10$, Algorithm~\ref{alg:randomcolumns} runs in
  time $\poly(nm)$ and returns $O(k \log m)$ columns that can be used
  as a factor of the whole matrix inducing $\ell_p$ error $O(k
  |\Delta|_p)$.
\end{theorem}
\begin{proof}
  First note, that if $|\Delta|_p \leq N \leq 2|\Delta|_p$ and if $i$
  is covered by a set $R$ of columns, then $i$ is $c_p$-approximately
  covered by $R$ for a constant $c_p$; here $c_p = 2^p$ for $p <
  \infty$ and $c_{\infty} = 2$.  By Lemma \ref{lem:cover}, the
  expected number of repetitions of selecting $2k$ columns until
  $(1/10)$-fraction of columns of $A$ are covered is $O(1)$. When we
  recurse on \SelectColumns on the resulting matrix
  $A_{\overline{R}}$, each such matrix admits a rank-$k$ factorization
  of cost at most $|\Delta|_p$.  Moreover, the number of recursive
  calls to \SelectColumns can be upper bounded by $\log_{10} m$. In
  expectation there will be $O(\log m)$ total repetitions of selecting
  $2k$ columns, and so by a Markov bound, with probability $9/10$, the
  algorithm will choose $O(k \log m)$ columns in total and run in time
  $\poly(nm)$.

  Let $S$ be the union of all columns of $A$ chosen by the
  algorithm. Then for each column $i$ of $A$, for $p \in [1, \infty)$,
  we have $\min_x |A_Sx-A_i|_p^p \leq \frac{100(k+1)^p 2^p
    |\Delta|_p^p}{n}$, and so $\min_X |A_SX-A|_p^p \leq 100(k+1)^p
  2^p |\Delta|_p^p$.  For $p = \infty$ we instead have $\min_x
  |A_Sx-A_i|_{\infty} \leq 2(k+1)|\Delta|_{\infty}$, and so $\min_X
  |A_SX-A|_{\infty} \leq 2(k+1)|\Delta|_{\infty}$.
\end{proof}

\subsection{A lower bound for \css{k}{p}}

In this section we prove an existential result showing that there
exists a matrix for which the best approximation to the \css{k}{p} is
$k^{\Omega(1)}$.

\begin{lemma}
  There exists a matrix $A$ such that the best approximation for the
  \css{k}{p} problem, for $p\in (2,\infty)$, is $k^{\Omega(1)}$.
\end{lemma}
\begin{proof}
  Consider $A = (k+1) I_{k+1}$, where $I_{k+1}$ is the $(k+1) \times (k+1)$
  identity matrix. And consider the matrix $B = (k+1) \cdot I_{k+1} - E$,
  where $E$ is the $(k+1) \times (k+1)$ all ones matrix. Note that $B$ has
  rank at most $k$, since the sum of its columns is $0$.

  {\bf Case: $2 < p < \infty$.}  If we choose any $k$ columns of $A$,
  then the $\ell_p$ cost of using them to approximate $A$ is
  $(k+1)$. On the other hand, $|A-B|_\infty = 1$, which means that
  $\ell_p$ cost of $B$ is smaller or equal than
  $\left((k+1)^2\right)^{1/p}$.
  
  {\bf Case: $p = \infty$.}  If we choose any $k$ columns of $A$, then
  the $\ell_\infty$ cost of using them to approximate $A$ is $k+1$. On
  the other hand, $|A-B|_\infty = 1$, which means that $\ell_\infty$
  cost of $B$ is smaller or equal than $1$.
\end{proof}
Note also that in~\cite{Woodruff2016} the authors show that for $p=1$
the best possible approximation is $\Omega(\sqrt{k})$ up to $\poly(\log k)$
factors.

%% file: iso.tex
\section{A $((k\log n)^k \poly(mn))$-time algorithm for \lra{k}{p}}
\label{sec:alg3}

In the previous section we have shown how to get a rank-$O(k\log m)$, 
$O(k)$-approximation in time $\poly(nm)$ to the \css{k}{p} and
\lra{k}{p} problems. In this section we first show how to get a rank-$k$,
$\poly(k)$-approximation efficiently starting from a rank-$O(k \log m)$ 
approximation.  This algorithm runs in polynomial time as long as $k =
O\left(\frac{\log n}{\log \log n}\right)$. We then show how to
obtain a $(k \log m)^{O(p)}$-approximation ratio in polynomial time
for every $k$. 

Let $U$ be the columns of
$A$ selected by Algorithm~\ref{alg:randomcolumns}.

\subsection{An isoperimetric transformation}

The first step of our proof is to show that we can modify the selected
columns of $A$ to span the same space but to have small distortion.
For this, we need the following notion of isoperimetry.
\begin{definition}[Almost isoperimetry]
  A matrix $B \in \reals^{n \times m}$ is
  \emph{almost-$\ell_p$-isoperimetric} if for all $x$, we have
  \mymath{ \frac{|x|_p}{2m} \leq |Bx|_p \leq |x|_p.  }
\end{definition}
We now show that given a full rank $A \in \reals^{n\times m}$, it is
possible to construct in polynomial time a matrix $B \in
\reals^{n\times m}$ such that $A$ and $B$ span the same space and $B$
is almost-$\ell_p$-isoperimetric.

\begin{lemma}\label{lem:iso}
  Given a full (column) rank $A \in \reals^{n\times m}$, there is an algorithm
  that transforms $A$ into a matrix $B$ such that $\spa A = \spa B$
  and $B$ is almost-$\ell_p$-isoperimetric. Furthermore the running
  time of the algorithm is
  $poly(nm)$. 
\end{lemma} 
\begin{proof}
  In~\cite{ddhkm07}, specifically, Equation (4) in the proof of
  Theorem 4, the authors show that in polynomial time it is possible
  to find a matrix $B$ such that $\spa B = \spa A$ and for all $x$,
  $$|x|_2 \leq |Bx|_p \leq \sqrt{m} |x|_2,$$ for any $p \geq 1$.

  If $p < 2$, their result implies 
  $$\frac{|x|_p}{\sqrt{m}} \leq |x|_2
  \leq |Bx|_p \leq \sqrt{m} |x|_2 \leq \sqrt{m} |x|_p,$$
   and so
  rescaling $B$ by $\sqrt{m}$ makes it almost-$\ell_p$-isoperimetric.
  On the other hand, if $p > 2$, then 
  $$|x|_p \leq |x|_2 \leq |Bx|_p
  \leq \sqrt{m} |x|_2 \leq m |x|_p,$$
   and rescaling $B$ by $m$ makes it 
  almost-$\ell_p$-isoperimetric.
\end{proof}
Note that the algorithm used in~\cite{ddhkm07} relies on the
construction of the L\"{o}wner--John ellipsoid for a specific set of
points.  Interestingly, we can also show that there is a more simple
and direct algorithm to compute such a matrix $B$; this may be of
independent interest.  We provide the details of our algorithm in the
supplementary material.

\subsection{Reducing the rank to $k$}

The main idea for reducing the rank is to first apply the
almost-$\ell_p$-isoperimetric transformation to the factor $U$ to
obtain a new factor $Z^0$. For such a $Z^0$, the $\ell_p$-norm of $Z^0
V$ is at most the $\ell_p$-norm of $V$. Using this fact we show that
$V$ has a low-rank approximation and a rank-$k$ approximation of $V$
translates into a good rank-$k$ approximation of $U V$. But a good
rank-$k$ approximation of $V$ can be obtained by exploring all
possible $k$-subsets of rows of $V$, as in
Algorithm~\ref{alg:nk}. More formally, in Algorithm~\ref{alg:red} we
give the pseudo-code to reduce the rank of our low-rank approximation
from $O(k \log m)$ to $k$. Let $\delta =
|\Delta|_p = \opt_{k,p}(A)$.

\begin{algorithm}
\begin{algorithmic}[1]
\REQUIRE $U \in \reals^{n\times O(k\log m)}$, $V \in \reals^{O(k\log m)\times m}$
\ENSURE $W \in \reals^{n\times k}$, $Z \in \reals^{k\times m}$
\STATE Apply Lemma~\ref{lem:iso} to $U$ to obtain matrix $W^0$
\STATE Apply Lemma~\ref{lem:secondlpfactor} to obtain matrix
$Z^0$, s.t. $\forall i$, $|W^0Z^0_i-(UV)_i|_p$  is minimized
\STATE Apply Algorithm~\ref{alg:nk} with input $(Z^0)^T \in
\reals^{n\times O(k\log m)}$ and $k$ to obtain $X$ and $Y$
\STATE Set $Z \leftarrow X^T$
\STATE Set $W \leftarrow W^0 Y^T$
\STATE Output $W$ and $Z$
\end{algorithmic}
\caption{\label{alg:red}An algorithm that transforms an $O(k\log m)$-rank matrix decomposition into a $k$-rank matrix decomposition without inflating the error too much.}
\end{algorithm}

\begin{theorem} \label{thm:polykApprox}
  Let $A \in \reals^{n\times m}$, $U \in \reals^{n\times O(k\log m)}$,
  $V \in \reals^{O(k\log m)\times m}$ be such that $|A - UV|_p =
  O(k\delta)$. Then, Algorithm~\ref{alg:red} runs in time
  $O(k\log m)^k(mn)^{O(1)}$ and outputs $W \in \reals^{n\times k},
  Z\in \reals^{k\times m}$ such that $|A - WZ|_p = O((k^{4}\log k) \delta)$. 
\end{theorem}
\iffull
\begin{proof}
  We start by bounding the running time.  Step $3$ is computationally
  the most expensive since it requires to execute a brute-force search
  on the $O(k\log m)$ columns of $(Z^0)^T$.  So the running time
  follows from Theorem~\ref{thm:nk}.

  Now we have to show that the algorithm returns a good approximation.
  The main idea behind the proof is that $UV$ is a low-rank
  approximable matrix. So after applying Lemma~\ref{lem:iso} to
  $U$ to obtain a low-rank approximation for $UV$ we can simply focus
  on $Z^0 \in \reals^{O(k\log m) \times n}$. Next, by applying
  Algorithm~\ref{alg:nk} to $Z^0$, we obtain a low-rank
  approximation in time $O(k\log m)^k(mn)^{O(1)}$. Finally we can use
  this solution to construct the solution to our initial problem.

  We know by assumption that $|A-UV|_p = O(k\delta)$.  Therefore, it
  suffices by the triangle inequality to show $|UV-WZ|_p = O((k^{4}\log
  k) \delta)$.  First note that $UV=W^0Z^0$ since Lemma~\ref{lem:iso}
  guarantees that $\spa U = \spa W^0$. Hence we can focus on proving
  $|W^0Z^0 - WZ|_p\leq O((k^{4}\log k) \delta)$.
  
  We first prove two useful intermediate steps.
\begin{lemma}
\label{lem:one}
 There exist matrices $U^*\in
    \reals^{n\times k}$, $V^*\in \reals^{k\times m}$ such that
    $|W^0Z^0-U^*V^*|_p = O(k\delta)$. 
\end{lemma}
\begin{proof} There exist $U^*\in \reals^{n\times k}$, $V^*\in
    \reals^{k\times m}$ such that $|A-U^*V^*|_p\leq \delta$ and,
    furthermore, $|A-UV|_p=|A-W^0Z^0|_p = O(k\delta)$.
    The claim follows by the Minkowski inequality. 
\end{proof}
\begin{lemma} 
There exist matrices $F \in \reals^{O(k \log m) \times
    k}$, $D \in \reals^{k \times n}$ such that $|W^0(Z^0- FD)|_p =
  O(k^{2}\delta )$. 
\end{lemma} 
\begin{proof} 
  From Lemma~\ref{lem:one}, we know that $|W^0Z^0-U^*V^*|_p =
  O(k\delta)$.  Hence, from Theorem~\ref{thm:kcol}, we know that there
  exists a matrix $C \in\reals^{n\times k}$ composed of $k$ columns of
  $W^0Z^0$, and a matrix $D \in \reals^{k\times m}$ such that
  $|W^0Z^0-CD|_p = O(k^{2}\delta)$.  Furthermore, note that selecting
  $k$ columns of $W^0Z^0$ is equivalent to select the same columns in
  $Z^0$ and multiplying them by $W^0$. So we can express $C= W^0F$ for
  some matrix $F\in \reals^{O(k\log m) \times k}$. Thus we can rewrite
\begin{eqnarray*}
|W^0Z^0-CD|_p & = & |W^0Z^0-W^0FD|_p \\
& = & |W^0(Z^0- FD)|_p \\
& \leq & O(k^{2}\delta).
\quad\qedhere
\end{eqnarray*}
\end{proof}
Now from the guarantees of Lemma~\ref{lem:iso} we know that for any
vector $y$, $|W^0 y|_p\leq \frac{|y|_p}{k\log k}$. So we have $|Z^0-
FD|_p\leq O((k^{3}\log k) \delta)$, Thus $|(Z^0)^T- D^TF^T|_p\leq
O((k^{3}\log k) \delta)$, so $(Z^0)^T$ has a low-rank approximation with
error at most $O((k^{3}\log k) \delta)$. So we can apply
Theorem~\ref{thm:kcol} again and we know that there are $k$ columns of
$(Z^0)^T$ such that the low-rank approximation obtained starting from
those columns has error at most $O((k^{4}\log k) \delta)$.  We obtain
such a low-rank approximation from Algorithm~\ref{alg:nk} with input
$(Z^0)^T \in \reals^{n\times O(k\log m)}$ and $k$. More precisely, we
obtain an $X\in \reals^{n\times k}$ and $Y \in \reals^{k\times O(k \log
  m)}$ such that $|(Z^0)^T- XY|_p\leq O((k^{4}\log k) \delta)$. Thus
$|Z^0- Y^TX^T|_p\leq O((k^{4}\log k) \delta)$.

Now using again the guarantees of Lemma~\ref{lem:iso} for $W^0$, we get
$|W^0(Z^0- Y^TX^T)|_p\leq O((k^{4}\log k) \delta)$. So
$|W^0(Z^0- Y^TX^T)|_p 
= |W^0Z^0- WZ)|_p
= |UV- WZ)|_p\leq O((k^{4}\log k) \delta)$. By
combining it with $|A-UV|_p = O(k\delta)$ and using the
Minkowski inequality, the proof is complete. \end{proof}\fi

\subsection{Improving the Running Time}
We now show how to improve the running time to $(mn)^{O(1)}$ for
every $k$ and every constant $p \geq 1$,
at the cost of a $\poly(k \log(m))$-approximation
instead of the $\poly(k)$-approximation we had previously.

\begin{theorem} 
  Let $A \in \reals^{n\times m}$, $1 \leq k \leq \min(m,n)$, and $p \geq 1$
  be an arbitrary constant. Let $U \in \reals^{n\times O(k\log m)}$ and 
  $V \in \reals^{O(k\log m)\times m}$ be such that $|A - UV|_p = O(k\delta)$.
  There is an algorithm which runs in time
  $(mn)^{O(1)}$ and outputs $W \in \reals^{n\times k},
  Z\in \reals^{k\times m}$ such that
  $|A - WZ|_p = (k \log m)^{O(p)} \delta$. 
\end{theorem}
\begin{proof}
  The proof of Theorem \ref{thm:polykApprox} shows there exists a rank-$k$
  matrix $X$ for which $|UXV^T-A|_p = O(k^4 \log k)\delta$. Instead
  of the enumeration algorithm used in the proof of Theorem
  \ref{thm:polykApprox} to find such an $X$, we will instead use
  $\ell_p$-leverage score sampling \cite{ddhkm07}.

  It is shown in Theorem 6 of 
  \cite{ddhkm07} that given $U$ one can in $\poly(mn)$ time and with
  probability $1-o(1)$, find a sampling
  and rescaling matrix $S$ with $(k \log m)^{O(p)}$ rows such that for
  all vectors
  $w$, $|SUw|_p = (1 \pm 1/2)|Uw|_p$. Indeed, in their notation
  one can compute a well-conditioned-basis in $\poly(mn)$ time and then
  sample rows of $U$ according to the $p$-th power of the
  $p$-norms of the rows of the
  well-conditioned basis. Since $S$ is a sampling and rescaling matrix,
  we have ${\bf E}[|SY|_p^p] = |Y|_p^p$ for any fixed matrix $Y$. 
  
  Let $X^*$ be the rank-$k$ matrix minimizing $|UX^*V^T-A|_p$. By the
  triangle inequality, for an arbitrary $X$ we have
  \begin{eqnarray}\label{eqn:related}
    |SUXV^T-SA|_p = |SUX^*V^T-SUXV^T|_p \pm |SUX^*V^T-SA|_p.
    \end{eqnarray}
  By a Markov bound, $|SUX^*V^T-SA|_p^p \leq 100^p |UX^*V^T-A|_p^p$
  with probability $1-1/100^p$, and so
  \begin{eqnarray}\label{eqn:Markov}
    |SUX^*V^T-SA|_p \leq 100 |UX^*V^T-A|_p
    \end{eqnarray}
  with this probability. Moreover, with probability $1-o(1)$,
  \begin{eqnarray}\label{eqn:subspace}
        |SUX^*V^T-SUXV^T|_p = (1 \pm 1/2)|UX^*V^T-UXV^T|_p
        \end{eqnarray}
  simultaneously for all $X$. By a union bound, both of these events occur with
  probability $1-1/100^p - o(1)$. In this case, it follows that if $X'$
  satisfies $|SUX'V^T-SA|_p \leq \alpha \min_{rank \ k \ B}|SUBV^T-SA|_p$,
  then also $|SUX'V^T-SA|_p \leq \alpha |SUX^*V^T-SA|_p$. Thus, using
  the triangle inequality, (\ref{eqn:Markov}) and (\ref{eqn:subspace}),
  \begin{eqnarray*}
    |UX'V^T-A|_p & \leq & |UX'V^T-UX^*V^T|_p + |UX^*V^T - A|_p\\
    & \leq & (1+1/2)|SUX'V^T-SUX^*V^T|_p + |UX^*V^T-A|_p\\
    & \leq & (1+1/2)(|SUX'V^T-SA|_p + |SUX^*V^T-SA|_p) + |UX^*V^T-A|_p\\
    & \leq & (1+1/2)((\alpha+1) |SUX^*V^T-SA|) + |UX^*V^T-A|_p\\
    & = & O(\alpha) |UX^*V^T-A|_p.
    \end{eqnarray*}

  Now consider the problem $|SUXV^T-SA|_p$. We can compute a
  well-conditioned basis in $\poly(mn)$ time and then sample columns of $V$
  according to the $p$-th power of the
  $p$-norms of the columns of the well-conditioned basis.
  Let $T$ denote this sampling matrix, which has $(k \log m)^{O(p)}$ columns.
  We condition on analogous events to those in (\ref{eqn:Markov}) and (\ref{eqn:subspace})
  above, which hold again with probability $1-1/100^p-o(1)$. Then 
  if $X''$ is a $\beta$-approximate minimizer to $|SUX''V^TT-SAT|_p$,
  then analogously,
  \begin{eqnarray}\label{eqn:2related}
  |SUX''V^T-SA|_p  \leq O(\beta)|SUX'V^T-SA|_p.
  \end{eqnarray}

  We thus have by several applications of the triangle inequality and the
  above, 
  \begin{eqnarray}\label{eqn:final}
    |UX''V^T-A|_p & \leq & |UX''V^T-UX'V^T|_p + |UX'V^T-A|_p \notag\\
    & \leq & (1+1/2)|SUX''V^T-SUX'V^T|_p + O(\alpha)|UX^*V^T-A|_p \notag \\
    & \leq & (1+1/2)(|SUX''V^T-SA|_p + |SUX'V^T-SA|_p) + O(\alpha \delta) \notag\\
    & \leq & O(\beta)|SUX'V^T-SA|_p + O(\alpha \delta) \notag \\
    & \leq & O(\beta)(|SUX'V^T-SUX^*V^T|_p + |SUX^*V^T-SA|_p) + O(\alpha \delta) \notag\\
    & \leq & O(\beta)(|UX'V^T-UX^*V^T|_p) + O((\alpha + \beta)\delta) \notag \\
    & \leq & O(\beta)(|UX'V^T-A|_p + \delta) + O((\alpha + \beta)\delta) \notag\\
    & \leq & O(\alpha \beta \delta).
    \end{eqnarray}

  Finally, observe that since $SUXV^TT-SAT$ is a $(k \log m)^{O(p)} \times (k \log m)^{O(p)}$ matrix for any $X$,
  it follows that its Frobenius norm is related up to a $(k \log m)^{O(p)}$ factor to its entrywise $p$-norm.
  Consequently, the Frobenius norm minimizer $X''$ is a $(k \log m)^{O(p)}$-approximate minimizer to the entrywise $p$-norm, and
  so $\beta = (k \log m)^{O(p)}$ in the notation above. It then follows from (\ref{eqn:2related}) that $\alpha = O(\beta) = (k \log m)^{O(p)}$
  as well. Consequently, by (\ref{eqn:final}), we have that $|UX''V^T-A|_p = (k \log m)^{O(p)} \delta$.

  Finally, note that the Frobenius norm minimizer $X''$ to $|SUX''V^TT-SAT|_p$ can be solved in time $(k \log m)^{O(p)}$ time, using the
  result in \cite{f07}. This completes the proof. 
  \end{proof}

%% file: exp.tex
\begin{figure*}[tp]%
    \centering
    \subfloat[FIDAP matrix ($\ell_1$)]{{\includegraphics[width=0.45\textwidth]{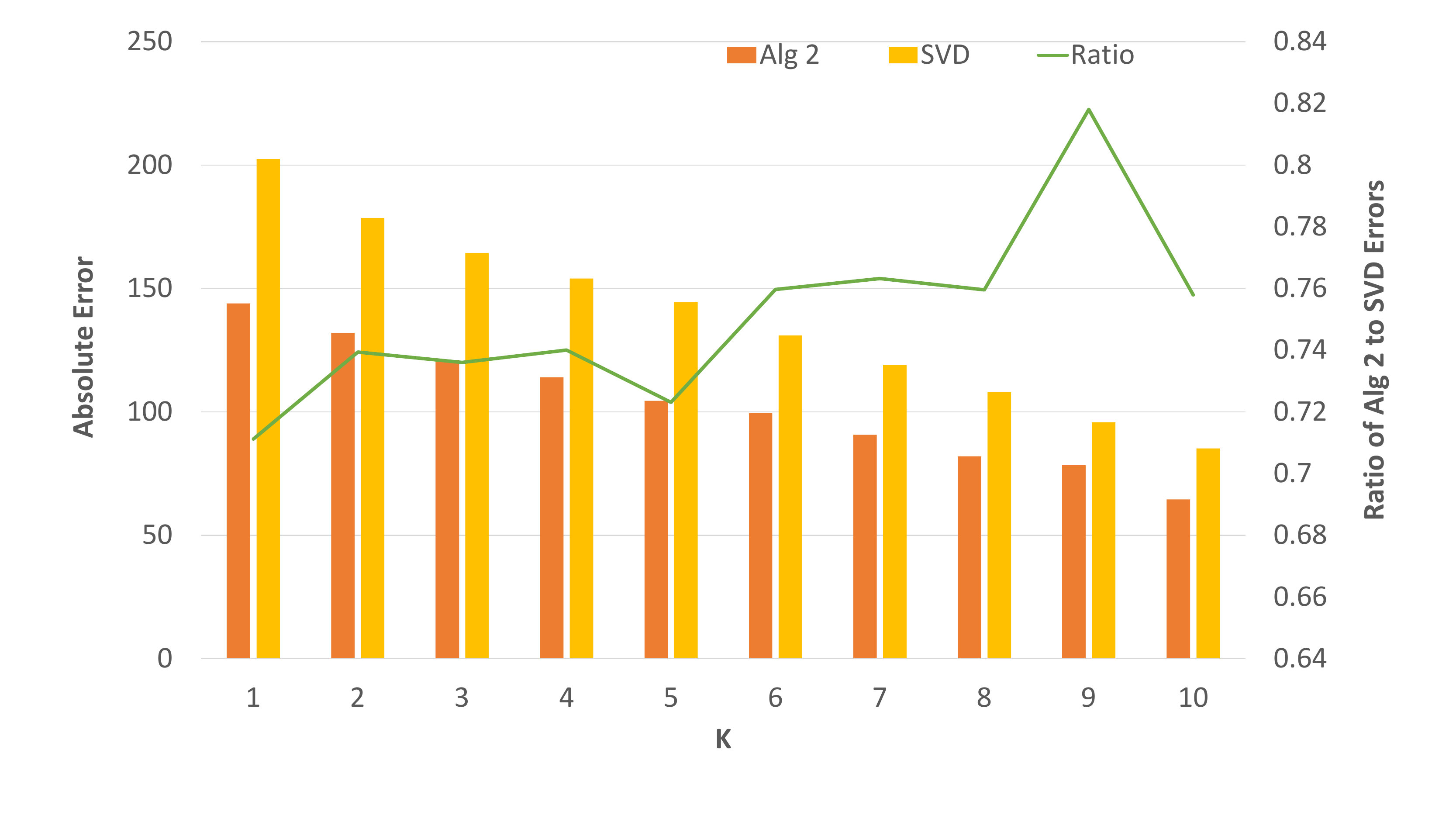} }}%
    \qquad
    \subfloat[KOS matrix ($\ell_\infty$)]{{\includegraphics[width=0.45\textwidth]{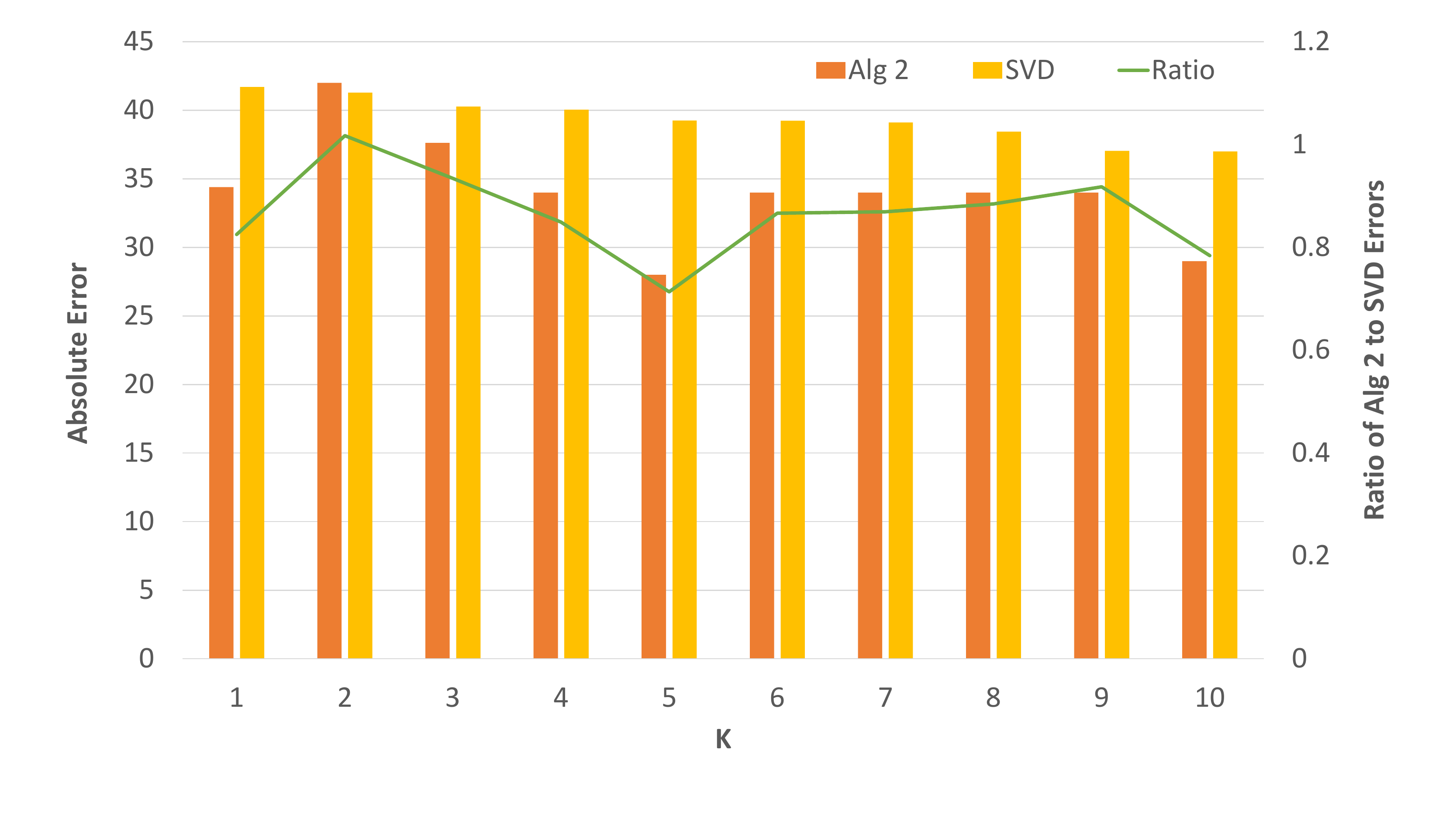} }}%
    \caption{Comparing the performance of Algorithm~\ref{alg:nk} with SVD on the real data sets.}%
    \label{fig:svd-cmp-real}%
\end{figure*}
\begin{figure*}[tp]%
    \centering
    \subfloat[Random matrix ($\ell_1$)]{{\includegraphics[width=0.45\textwidth]{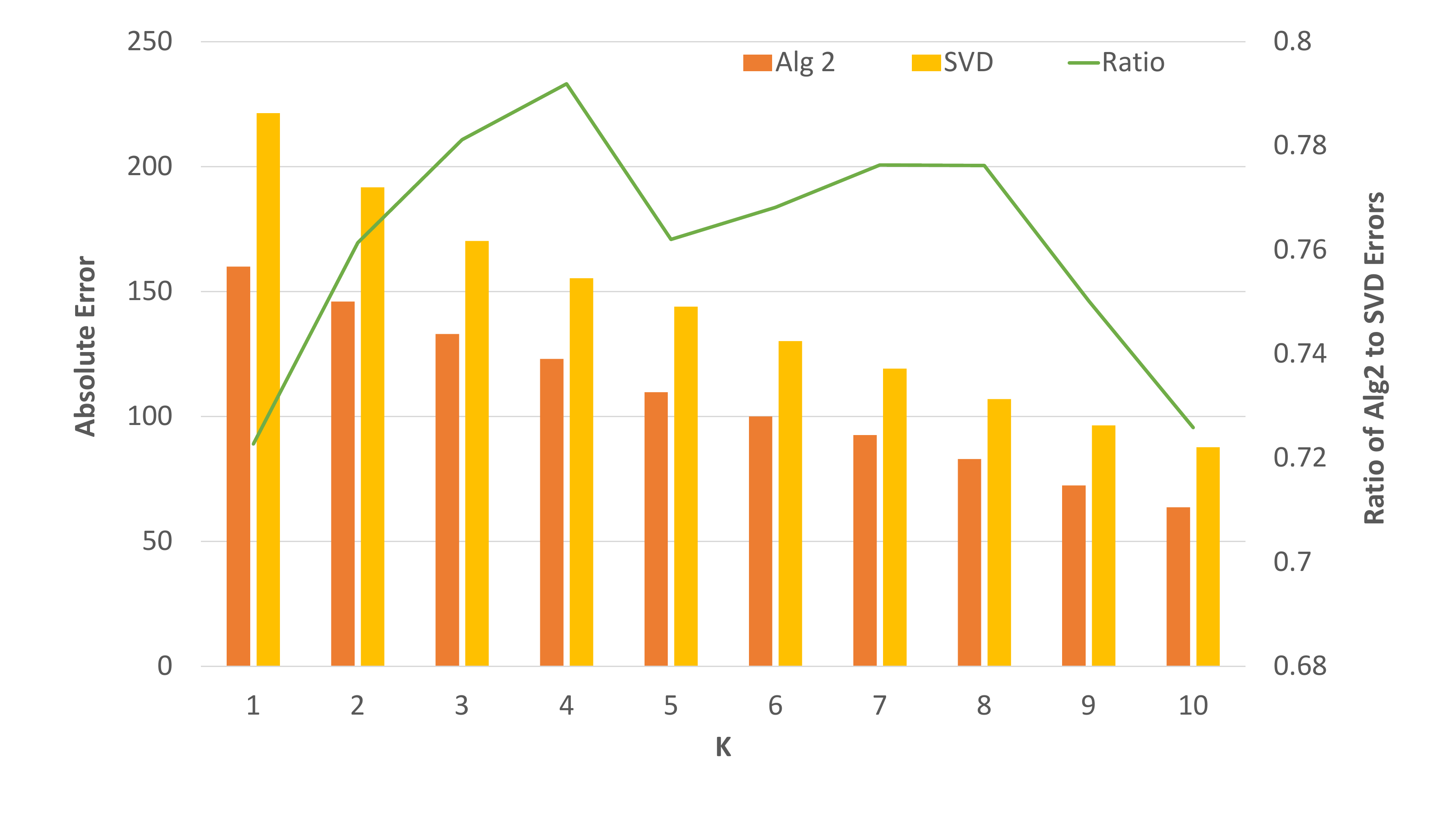} }}%
    \qquad
    \subfloat[Random matrix ($\ell_\infty$)]{{\includegraphics[width=0.45\textwidth]{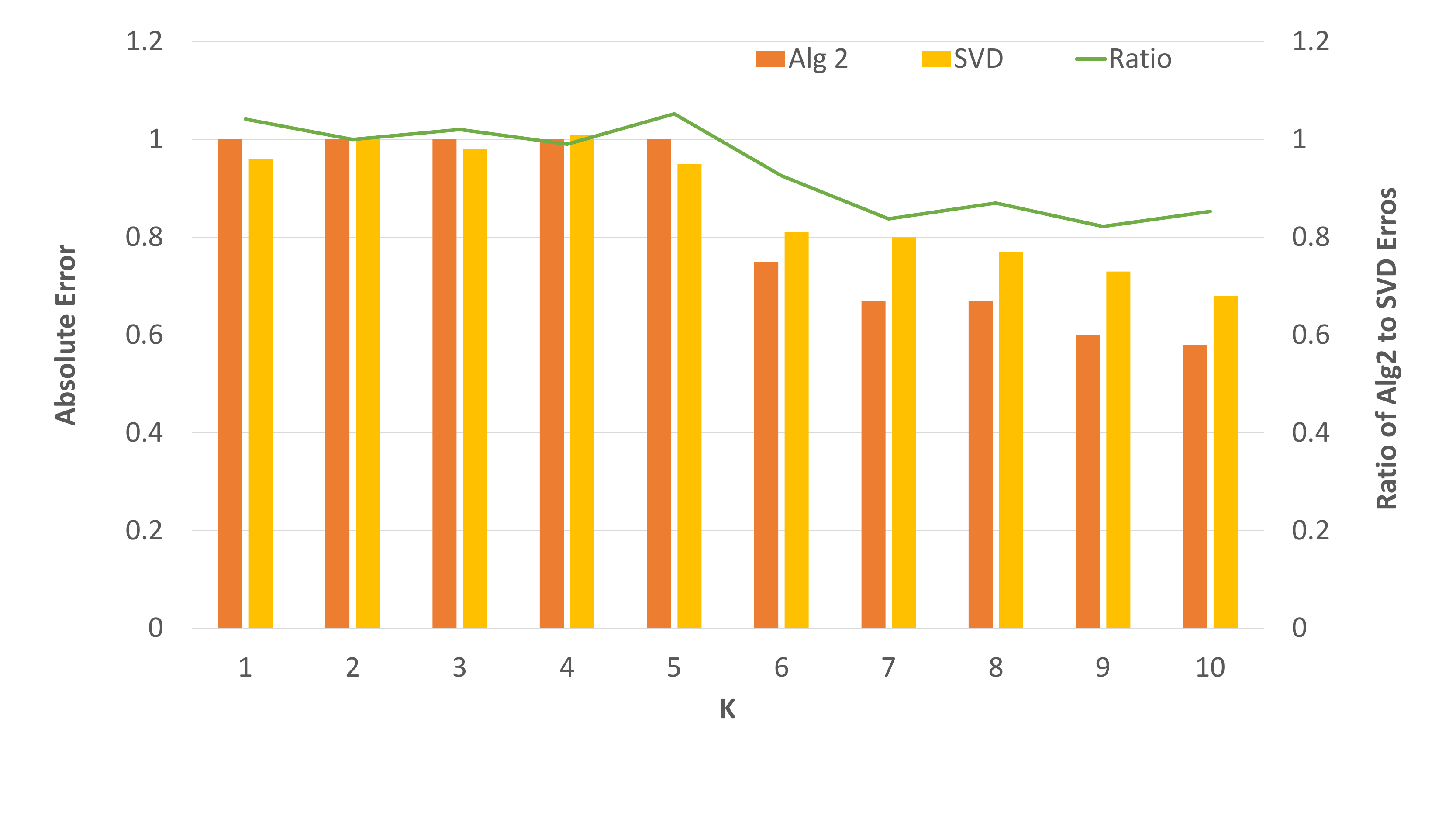} }}%
    \caption{Comparing the performance of Algorithm~\ref{alg:nk} with SVD on the random matrix.}%
    \label{fig:svd-cmp-random}%
\end{figure*}
\begin{figure*}[tp]%
    \centering
    \subfloat[$\pm 1$ matrix ($\ell_1$)]{{\includegraphics[width=0.45\textwidth]{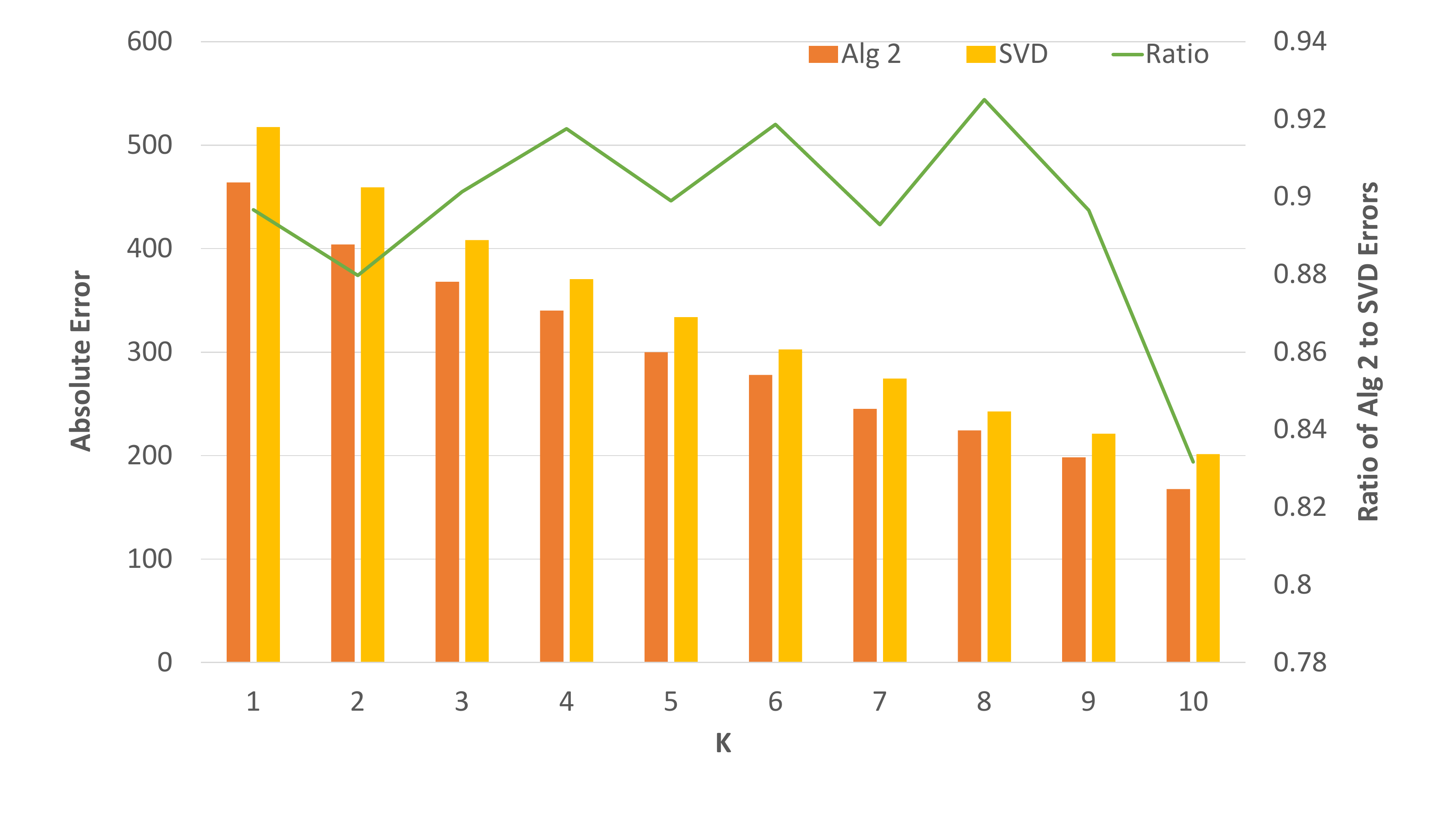} }}%
    \qquad
    \subfloat[$\pm 1$ matrix ($\ell_\infty$)]{{\includegraphics[width=0.45\textwidth]{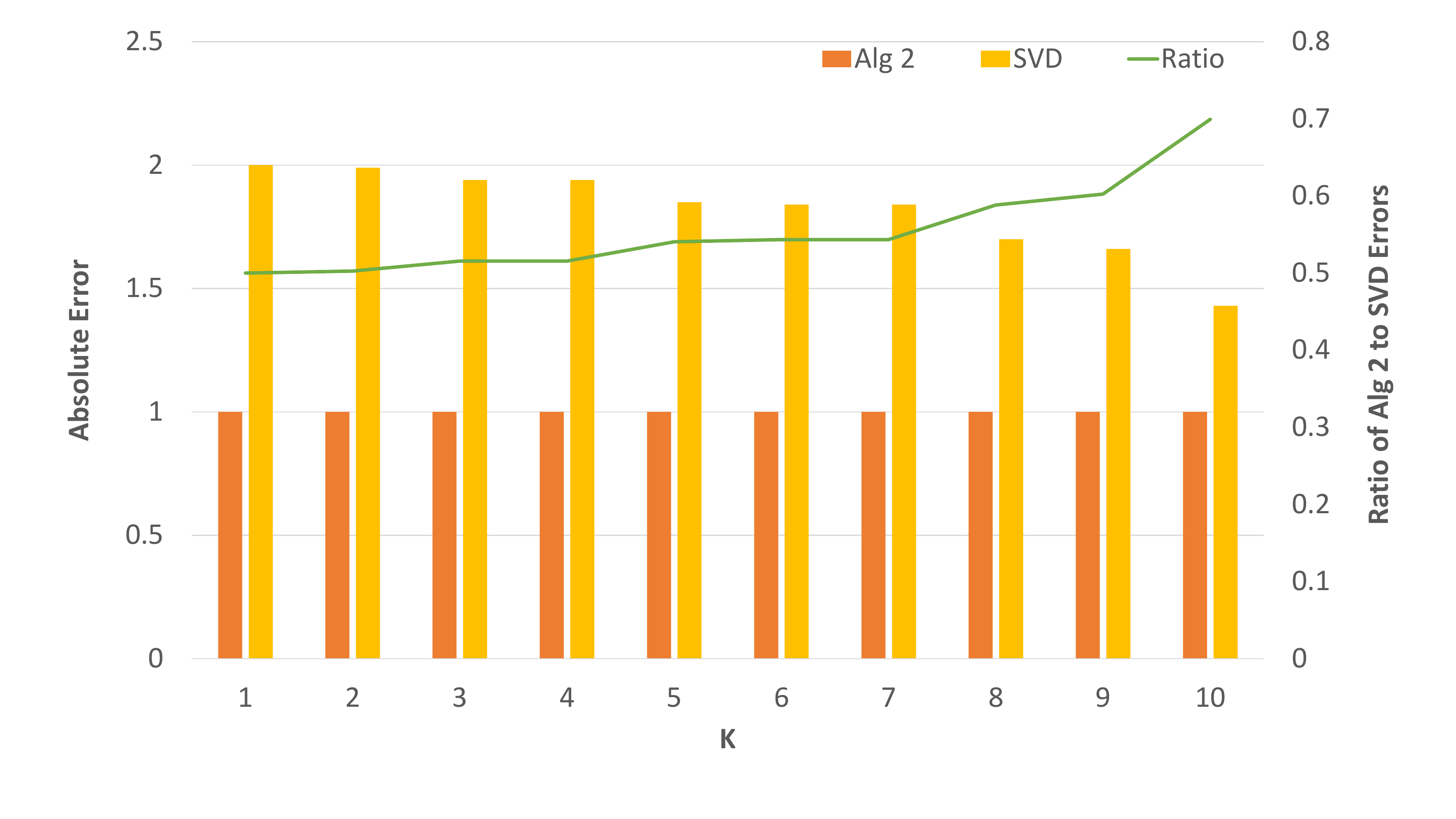} }}%
    \caption{Comparing the performance of Algorithm~\ref{alg:nk} with SVD on the $\pm 1$ matrix.}%
    \label{fig:svd-cmp-pm}%
\end{figure*}

\section{Experiments}
\label{sec:expt}

In this section, we show the effectiveness of Algorithm~\ref{alg:nk}
compared to the SVD. We run our comparison both on synthetic as well
as real data sets. For the real data sets, we use matrices from the
FIDAP
set\footnote{i\url{http://math.nist.gov/MatrixMarket/data/SPARSKIT/fidap/fidap005.html}}
and a word frequency dataset from UC
Irvine~\footnote{\url{https://archive.ics.uci.edu/
    ml/datasets/Bag+of+Words}}. The FIDAP matrix is 27 $\times$ 27
with 279 real asymmetric non-zero entries. The KOS blog entries
matrix, representing word frequencies in blogs, is 3430 $\times$ 6906
with 353160 non-zero entries.
For the synthetic data sets, we use two matrices. For the first, we
use a 20 $\times$ 30 random matrix with 184 non-zero entries---this
random matrix was generated as follows: independently, we set each
entry to $0$ with probability $0.7$, and to a uniformly random value
in $[0,1]$ with probability $0.3$. Both matrices are full rank. For
the second matrix, we use a random $\pm 1$ 20 $\times$ 30 matrix.

In all our experiments, we run a simplified version of
Algorithm~\ref{alg:nk}, where instead of running for all possible
$\binom{[m]}k$ subsets of $k$ columns (which would be computationally
prohibitive), we repeatedly sample $k$ columns, a few thousand times,
uniformly at random. We then run the $\ell_p$-projection on each
sampled set and finally select the solution with the smallest
$\ell_p$-error. (While this may not guarantee provable approximations,
we use this a reasonable heuristic that seems to work well in
practice, without much computational overhead.) We focus on $p = 1$
and $p = \infty$.

Figure~\ref{fig:svd-cmp-real} illustrates the relative performance of
Algorithm~\ref{alg:nk} compared to the SVD for different values of $k$
on the real data sets. In the figure the green line is the ratio
of the total error. The $\ell_1$-error for Algorithm~\ref{alg:nk}
is always less than the corresponding error for the SVD and in fact
consistently outperforms the SVD by roughly 40\% for small values of
$k$ on the FIDAP matrix. On the larger KOS matrix, the relative
improvement in performance with respect to $\ell_{\infty}$-error is
more uniform (around 10\%).  

We observe similar trends for the synthetic data sets as well.
Figures~\ref{fig:svd-cmp-random} and~\ref{fig:svd-cmp-pm} illustrate
the trends. Algorithm~\ref{alg:nk} performs consistently better than
the SVD in the case of $\ell_1$-error for both the matrices. In the
case of $\ell_\infty$-error, it outperforms SVD by around $10\%$ for
higher values of $k$ on the random matrix. Furthermore, it
consistently outperforms SVD, between 30\% and 50\%, for all values of
$k$ on the random $\pm 1$ matrix.

To see why our $\ell_{\infty}$ error is always $1$ for a random $\pm 1$ matrix $A$, note that by setting our rank-$k$ approximation to be the zero matrix, we achieve an $\ell_{\infty}$ error of $1$. This is optimal for large values of $n$ and $m$ and small $k$ as can be seen by recalling the notion of the {\it sign-rank} of a matrix $A \in \{-1,1\}^{n \times m}$, which is the minimum rank of a matrix $B$ for which the sign of $B_{i,j}$ equals $A_{i,j}$ for all entries $i,j$. If the sign-rank of $A$ is larger than $k$, then for any rank-$k$ matrix $B$, we have $\|A-B\|_{\infty} \geq 1$ since necessarily there is an entry $A_{i,j}$ for which $|A_{i,j} - B_{i,j}| \geq 1$. It is known that the sign-rank of a random $m \times m$ matrix $A$, and thus also of a random $n \times m$ matrix $A$, is $\Omega(\sqrt{m})$ with high probability \cite{f02}. 

%% file: conc.tex
\section{Conclusions}

We studied the problem of low-rank approximation in the entrywise $\ell_p$
error norm and obtained the first provably good approximation algorithms for the
problem that work for every $p \geq 1$.
Our algorithms are extremely simple, which
makes them practically appealing.
We showed the effectiveness of our algorithms
compared with the SVD on real and synthetic data sets. 
We obtain a $k^{O(1)}$ approximation factor for every $p$ for the column
subset selection problem, and we 
showed an example matrix for this problem
for which a $k^{\Omega(1)}$ approximation factor is necessary. 
It is unclear if better approximation
factors are possible by designing algorithms that do not choose
a subset of input columns to span the output low rank
approximation.  Resolving this
would be an interesting and important research direction.